\documentclass[%
 reprint,
 amsmath,amssymb,
 aps,
prl,
superscriptaddress
]{revtex4-2}

\usepackage{graphicx}% Include figure files
\usepackage{dcolumn}% Align table columns on decimal point
\usepackage{bm}% bold math
\usepackage[colorlinks=true, allcolors=blue]{hyperref}
\usepackage{braket}
\usepackage{bbding}
\usepackage{natbib}
\usepackage{gensymb}
\usepackage{balance}
\usepackage{makecell}
\begin{document}

\preprint{APS/123}

\title{Magnon Nonlinear Hall Effect in 2D Antiferromagnetic Insulators} %: \\
%External-field Manipulation and Layer Dependence}% Force line breaks with %\\
%\thanks{A footnote to the article title}%

\author{Jinyang Ni}
\email{jinyang.ni@ntu.edu.sg}
\affiliation{%
 Division of Physics and Applied Physics, School of Physics and Mathematical Sciences, Nanyang Technological University, Signapore 637731, Singapore %\textbackslash\textbackslash
}%
% \altaffiliation{SPMS, Nanyang Techonology University.}%Lines break automatically or can be forced with \\
\author{Yuanjun Jin}
\affiliation{%
 Guangdong Basic Research Center of Excellence for Structure and Fundamental Interactions of Matter,
 Guangdong Provincial Key Laboratory of Quantum Engineering and Quantum Materials,
 School of Physics, South China Normal University, Guangzhou 510006, China %\textbackslash\textbackslash
}%
\author{Quanchao Du}
%\email{jinyang.ni@ntu.edu.sg}
\affiliation{%
 Division of Physics and Applied Physics, School of Physics and Mathematical Sciences, Nanyang Technological University, Signapore 637731, Singapore %\textbackslash\textbackslash
}%
\author{Guoqing Chang}
\email{guoqing.chang@ntu.edu.sg}
\affiliation{%
 Division of Physics and Applied Physics, School of Physics and Mathematical Sciences, Nanyang Technological University, Signapore 637731, Singapore %\textbackslash\textbackslash
}%

%\collaboration{MUSO Collaboration}%\noaffiliation

%\author{Charlie Author}
 %\homepage{http://www.Second.institution.edu/~Charlie.Author}
%\affiliation{
% Second institution and/or address\\
% This line break forced% with \\
%}%
%\affiliation{
% Third institution, the second for Charlie Author
%}%
%\author{Delta Author}
%\affiliation{%
% Authors' institution and/or address\\
% This line break forced with \textbackslash\textbackslash
%}%

%\collaboration{CLEO Collaboration}%\noaffiliation

%\date{\today}% It is always \today, today,
             %  but any date may be explicitly specified

\begin{abstract}
The efficient detection of the magnetism in 2D antiferromagnetic\,(AFM) insulators is crucial for the advancement of 2D AFM spintronics and remains a challenging problem. In this letter, we introduce the magnon nonlinear Hall current, a second-order Hall response of collective spin excitations in ordered magnets, as a novel probe for 2D layered AFM insulators. We theoretically demonstrate that the nonlinear Hall effect is intrinsically coupled to the underlying spin configuration. In particular, it exhibits a pronounced layer dependence in layered antiferromagnets, enabling direct characterization of the nature and strength of interlayer magnetic coupling in multilayer AFM insulators. Furthermore, we show that a a slight external field perturbation can induce and manipulate the magnon nonlinear Hall response. Our work establishes a novel approach for exploring 2D antiferromagnetism and holds great promise for AFM spintronic applications. 
%\begin{description}
%\item[Usage]
%xxx.
%\item[Structure]
%xxx. 
%\end{description}
\end{abstract}

%\keywords{Suggested keywords}%Use showkeys class option if keyword
                              %display desired
\maketitle

%\tableofcontents

%\section{Introduction}
\textit{Introduction.} Two-dimensional (2D) van der Waals (vdW) magnets provide an ideal platform for fundamental studies of magnetism and magnetic excitations in the 2D limit\,\cite{burch2018magnetism, mak2019probing, rodin2020collective, cenker2021direct, bae2022exciton}. Their highly tunable magnetism, coupled to the
ability to form stacked heterostructures, enable access to non-trivial magnetic phases and various technological applications, including compact memory storage devices, sensors and quantum computing\,\cite{burch2018magnetism, mak2019probing, rodin2020collective, cenker2021direct, gong2019two, gibertini2019magnetic, bae2022exciton,burch2018magnetism, liu20192d}. In addition to ferromagnetism, first observed in $\mbox{CrI}_{3}$\,\cite{huang2017layer} and $\mbox{CrGeTe}_{3}$\,\cite{gong2017discovery}, antiferromagnetism in 2D materials, exemplified by the family ${M\mbox{P}X_{3}}$  ($M=\mbox{V},\mbox{Mn},\mbox{Fe},\mbox{Co},\mbox{Ni};X=\mbox{S},\mbox{Se}, \mbox{Te}$)\,\cite{du2016weak,wang2018new, ni2021direct, kim2019suppression, klaproth2023origin, ni2021imaging, liu2023probing}, has attracted much attention due to its promising applications. Compared to ferromagnets, antiferromagnets produce no stray field, display ultrafast dynamics in the terahertz (THz) regime, and demonstrate robustness against external field perturbation\,\cite{neel1948proprietes, jungwirth2016antiferromagnetic, sano2024acoustomagnonic, baltz2018antiferromagnetic}. Thus, the study of 2D antiferromagnets not only enhances fundamental understanding low-dimensional magnetism, but also paves new avenues for the realization of high speed, compact spintronic and magnonic devices\,\cite{baltz2018antiferromagnetic, chumak2015magnon}.

However, current experimental approaches for detecting atomic-thin antiferromagnetic (AFM) insulators face several significant limitations. Traditional techniques, such as magnetization measurements and neutron diffraction, are primarily designed to probe macroscopic magnetic properties and are thus unsuitable for studying atomically thin magnetic materials\,\cite{gibertini2019magnetic, mak2019probing, huang2020emergent}. Alternative methods designed for 2D systems often rely on electrical conductivity, specific optical modes, or ferromagnetic ordering, making their application to 2D AFM insulators challenging\,\cite{mak2019probing, houmes2023magnetic}. Furthermore, the weak vdW interactions enables diverse stacking configurations, introducing complexity to the study of 2D antiferromagnetism\,\cite{chu2020linear, sivadas2018stacking, jiang2019stacking}. Therefore, a comprehensive method capable of exploring the magnetic properties of 2D layered AFM insulators is highly desired.

\begin{figure} 
\includegraphics[width=0.45\textwidth]{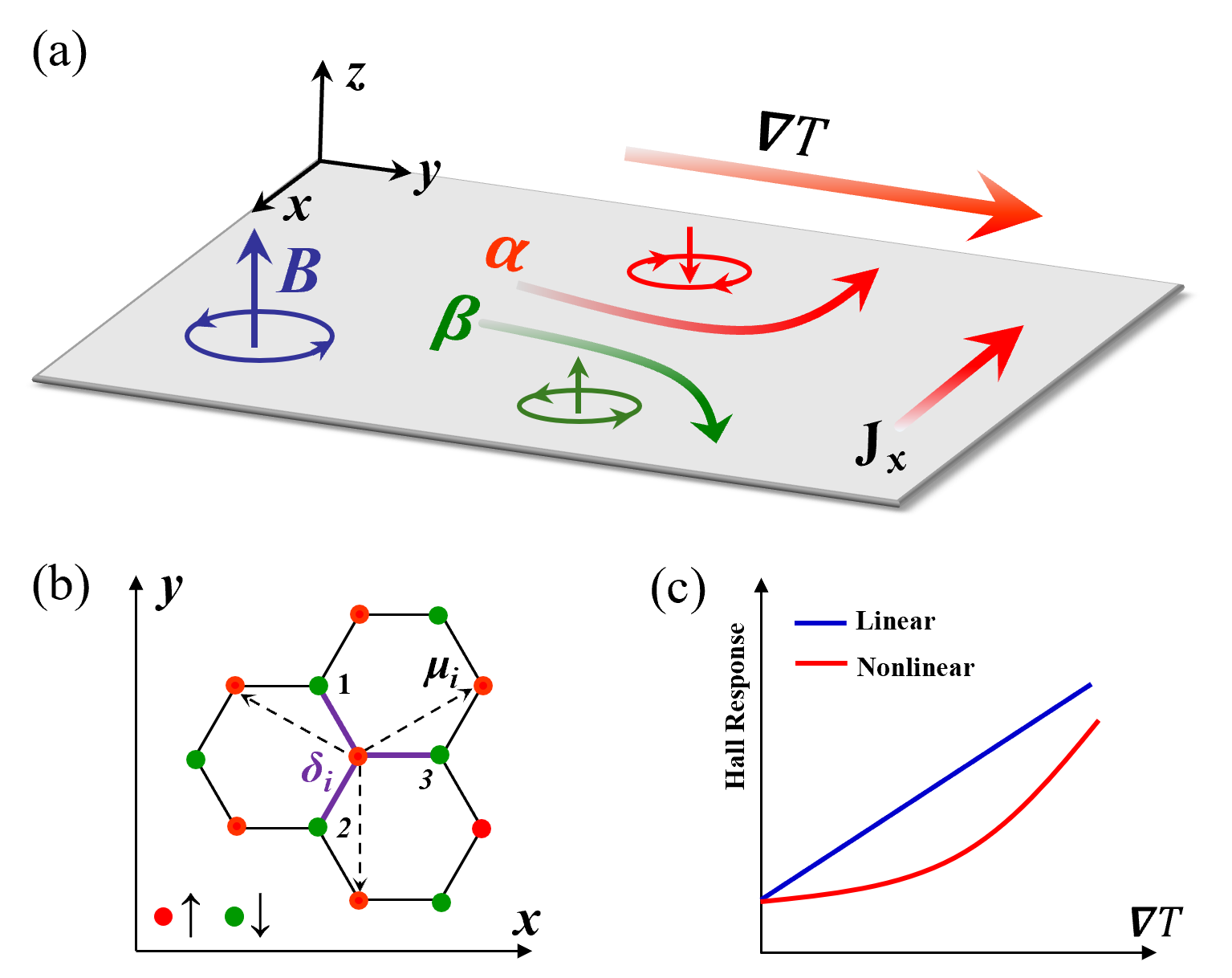}% Here is how to import EPS art
\caption{\label{fig1}(a) Schematic illustration of magnon nonlinear Hall effect in antiferromagnets. (b) Lattice structure of honeycomb antiferromagnet with N\'{e}el order, where $\delta_{i}$ and $\mu_{i}$ connects NN and 2NN sites, respectively. (c) A comparison between
the linear (red line) and nonlinear (green line) magnon Hall effect.}
\end{figure}

In this letter, we unveil an intriguing magnetic-layer dependence of magnon nonlinear Hall response\,(MNHR), which can be used as a novel approach to probe the magnetic structures in 2D AFM insulators. As schematically illustrated in Fig.\,\ref{fig1}(a), the nonlinear Hall current of magnons can be induced by applying a finite external-field ${\cal B}$ in layered AFM insulators, in contrast to the known nonlinear Hall effects observed in phonons or electrons\,\cite{sodemann2015quantum, luo2023nonlinear}. Due to the magnon chirality in antiferromagnets is coupled to the $\cal {B}$, the sign of MNHE can be reversed by flipping ${\cal B}$. Therefore, the MNHE, carrying spin chirality information, presents a promising concept for detection and manipulation of 2D antiferromagnetism. Additionally, the coupling to the magnetic order give rise an intriguing layer dependence for MNHR, offering an effective method to characterize the nature and strength of interlayer coupling in multilayer antiferromagnets. By performing density-functional-theory\,(DFT) calculations, we validate the magnetic-layer dependence of MNHR in layered vdW AFM insulators ${\mbox{VP}X_{3}}$.

\textit{The origin of MNHR.} We firstly examine the origin of the MNHE in monolayer honeycomb antiferromagnet\,($m$AFM) with N\'{e}el order, where the minimal spin model can be expressed as:
\begin{equation} \label{Ham_1}
\begin{split}
{\cal \hat{H}}_{m}  = & {\cal J}_{1} \sum_{\langle {i,j} \rangle}{{\cal S}_{i} \cdot{\cal S}_{j}} + {\cal D}_{z} \sum_{ \langle {i,j} \rangle^{\prime}} \boldsymbol{\nu}_{ij}{ \cdot \left( {\cal S}_{i} \times {\cal S}_{j} \right)}  \\ & + \sum_{i} {\cal K} \left( {\cal S}^{z}_{i} \right)^{2} - {\cal B}\sum_{i}{\cal S}^{z}_{i}. 
\end{split}
\end{equation}
Here, the first term is Heisenberg exchange interaction with ${\cal J}_{1}$\,$>$\,$0$ for nearest-neighbor (NN) magnetic moments on honeycomb lattice. The second term is next-nearest-neighbor (2NN) out-of-plane Dzyaloshinskii-Moriya interaction (DMI) ${\cal D}_{z}$ \,\cite{dzyaloshinsky_JPCS_1958_4, moriya_PR_1960_120, mook_PRX_2021_11}, where the direction is defined as $\boldsymbol{\nu}_{ij}= 2\sqrt{3}\boldsymbol{\delta}_{i}\times\boldsymbol{\delta}_{j} = \pm{\boldsymbol{z}}$, with $\boldsymbol{\delta}_{i}$ being NN linking vectors, as shown in Fig.\,\,\ref{fig1}(b). The third term represents the easy-axis single ion anisotropy (SIA) with ${\cal K}$\,$<$\,$0$. For simplicity, the external Zeeman field $\cal B$ is applied along $z$-axis. Notice that the following discussion is restricted to $|{\cal J}_{1}|$\,$\gg$\,$|{\cal D}_{z}|$ and $2|{\cal K}|S$\,$>$\,$|{\cal B}|$, ensuring that the magnetic ground state remains in the collinear N\'{e}el order. 
\begin{figure}
\includegraphics[width=0.475\textwidth]{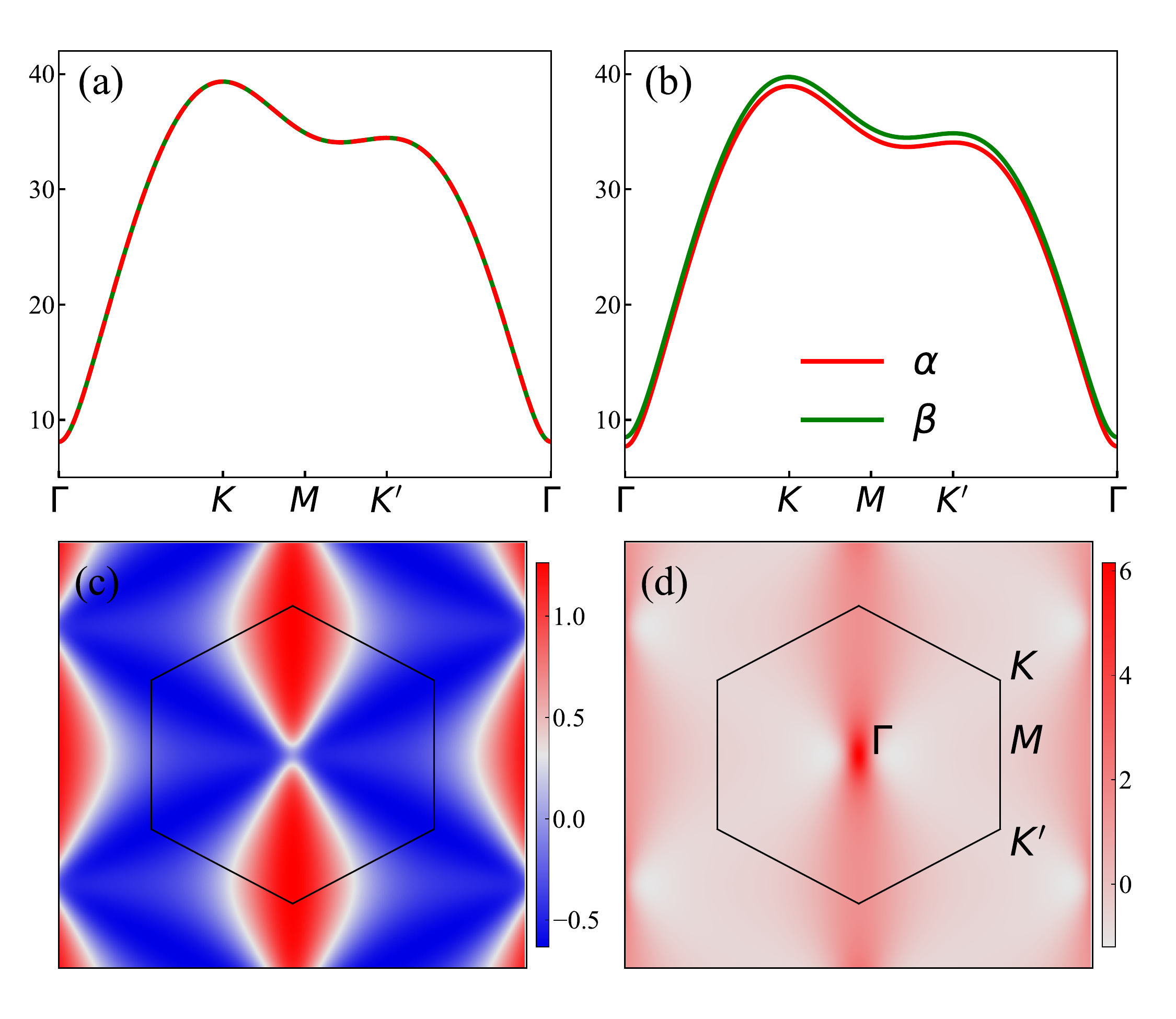}% Here is how to import EPS art
\caption{\label{fig2}(a)-(b) Magnon bands (units of meV) of monolayer ${\mbox{VPTe}_{3}}$ with (a) ${\cal B}$\,$=$\,$0$ and (b) ${\cal B}$\,$=$\,$-0.4$\,$\mbox{meV}$. Clearly, the bands both gap at $\Gamma$-point with the gap being equal to $S(2\sqrt{|{\cal K}|^{2} - 3{\cal KJ}_{1}} \pm {\cal B})$. (c) Corresponding extended BCD of magnons (minor band) without strain, (d) with $3\%$ compressive strain.}
\end{figure}

The linear spin wave model in Eq.\,(\ref{Ham_1}) can be solved by employing Holstein-Primakoff (HP) transformation\,\cite{holstein1940field}, ${\cal S}^{+}_{i,\uparrow}$\,$\approx$\, $\sqrt{2S}\hat{a}_{i}$, ${\cal S}^{+}_{i,\downarrow}$\,$\approx$\, $\sqrt{2S}\hat{b}^{\dagger}_{i}$, ${\cal S}^{z}_{i,\uparrow}$\,$=$\,$S$\,$-$\,$\hat{a}_{i}^{\dagger}\hat{a}_{i}$ and ${\cal S}^{z}_{i,\downarrow}$\,$=$\,$\hat{b}_{i}^{\dagger}\hat{b}_{i}$\,$-S$. Upon Fourier transformation, the Hamiltonian can be expressed in the spinor basis $\psi^{\dagger}_{k} = (\hat{a}_{k}^{\dagger}, \hat{b}_{-k}, \hat{a}_{-k}, \hat{b}^{\dagger}_{k})$ as ${\cal \hat{H}}_{m} = \sum_{k}\psi^{\dagger}_{k} {\cal \hat{H}}_{mk} \psi_{k}$. Neglecting the zero-point energy and assuming $S$\,$=$\,$1$, ${\cal \hat{H}}_{k}$ reads as, 
%\begin{widetext}
\begin{equation} {\label{Ham_k}}
 {\cal \hat{H}}_{mk} =  \lambda I + \left(\begin{matrix}
    {f_{k}} + {\cal B}&\gamma_{k} & 0 & 0\\
    \gamma_{k}^{\dagger}&  -  {f_{k}}-{\cal B} &0 & 0 \\
    0 & 0 & {\cal B} - {f_{k}} &  \gamma_{k}^{\dagger} \\
    0 & 0 &  \gamma_{k} &  {f_{k}} - {\cal B}
\end{matrix} \right),
\end{equation}
%\end{widetext}}
where $\lambda$\,$=$\,$-3{\cal J}_{1}$\,$-$\,$2{\cal K}$,
$\gamma_{k}$\,$=$\,$ {\cal J}_{1}\sum_{i} \mbox{exp}(i\textbf{\textit{k}}\cdot\boldsymbol{\delta}_{i})$ and $f_{k}$\,$=$\,${\cal D}_{z}\sum_{i\in odd}2\mbox{sin}(\textbf{\textit{k}}\cdot \boldsymbol{\mu}_{i})$ with $\boldsymbol{\mu}_{i}$ the vectors linking 2NN sites as shown in Fig.\,\ref{fig1}(b).

Using Bogoliubov transformation\,\cite{bogoljubov1958new, valatin1958comments}, we can derive the analytical eigenvalues and eigenvectors of Eq.\,(\ref{Ham_k}), expressed as
\begin{equation}\label{eigen_AFM}
     {{\epsilon}}_{\pm} = \epsilon_{0} + f_{k} \pm {\cal B}, \
     \Psi^{\dagger}_{\pm} = \frac{\sqrt{2}}{2}\left( -\frac{\lambda \pm \epsilon_{0}}{\gamma_{k}(\gamma^{\dagger}_{k})},  1\right), 
\end{equation}
where $\pm$ represent the $\alpha$ and $\beta$ mode, respectively, and $\epsilon_{0} = \sqrt{\lambda^{2}-|\gamma_{k}|^{2}}$. The corresponding Berry curvature can therefore be determined as follows
\begin{equation}
    \boldsymbol{\Omega}^{z}_{\pm} (\boldsymbol{k}) = \pm \frac{\lambda}{2\epsilon_{0}^{3}} \left (\boldsymbol{\nabla}{\mbox{Re}\gamma_{k}} \times \boldsymbol{\nabla}{\mbox{Im}\gamma_{k}}  \right).
\end{equation}
Clearly, the Berry curvature is independent of ${\cal D}_{z}$ and ${\cal B}$. 
This can be explained by the symmetry argument of the N\'{e}el order\,\cite{brinkman1966theory}, which naturally breaks the inversion symmetry ${\cal P}$, leading to $\boldsymbol{\Omega}(k) = -\boldsymbol{\Omega}(-k)$. Thus, the nonzero ${\boldsymbol{\Omega}}$ can arise even when both ${\cal B}$ and ${\cal D}_{z}$ are zero. In this scenario, two degenerate magnon modes of $m$AFM, $\alpha$ and $\beta$, are symmetric bands about $\Gamma$ point, characterized by spin angular value $\langle S^{z} \rangle$ with chirality $\mp{1}$.

\begin{table}[b]
\caption{\label{tab_1}\textcolor{red}{The possible magnetic point group support to generate the nonlinear magnons Hall response in 2D limit.}}
\begin{ruledtabular}
\begin{tabular}{@{}c >{\centering\arraybackslash}p{0.75\linewidth}@{}}
Bravais lattice & Magnetic point group  \tabularnewline
\hline 
\mbox{Oblique}  &  $\bar{1}^{\prime}$ \tabularnewline
\mbox{Rectangular} & $2^{\prime}/m$, $2/m^{\prime}$, 
$m^{\prime}m^{\prime}m^{\prime}$, $mmm^{\prime}$  \tabularnewline
\mbox{Square} & $4/m^{\prime}$, $4^{\prime}/m^{\prime}$, $4/m^{\prime}mm$, $4^{\prime}/m^{\prime}m^{\prime}m$, $4/m^{\prime}m^{\prime}m^{\prime}$ \tabularnewline 
\mbox{Hexagonal} & $\bar{3}^{\prime}$, $\bar{3}^{\prime}m$, $\bar{3}^{\prime}m^{\prime}$, 
$6^{\prime}/m$, $6/m^{\prime}$, $6^{\prime}/mmm^{\prime}$, $6/m^{\prime}mm$, $6/m^{\prime}m^{\prime}m^{\prime}$
\end{tabular}
\end{ruledtabular}

\end{table} 

As shown in Fig. \ref{fig2}(a), the ${\cal D}_{z}$ breaks the combined symmetry $C_{2}{\cal T}$\,(${\cal T}^{\prime}$), where $C_{2}$ denotes the $\pi$ rotation along $x/y$-axis and ${\cal T}$ is time-reversal symmetry, resulting in $\epsilon(k)$\,$\neq$\,$\epsilon(-k)$. This correction can induce the emergence of spin-Hall like magnon current by introducing the temperature gradient $\boldsymbol{\nabla}{T}$, known as linear magnon spin Nernst effect\,(SNE) \,\cite{cheng2016spin, zyuzin2016magnon, lee_prb_2018_97}. In addition to linear Hall response, the odd nature of $\boldsymbol{\Omega}(k)$, arising from the broken inversion symmetry $\cal P$, gives rise to a nonlinear Hall response\,\cite{sodemann2015quantum}. By solving the Boltzmann equation, as
detailed in supplemental materials(SM)\,\cite{Supplemental_Materials}, we derived the nonlinear spin-Nernst conductivity $I^{s}_{2}$ and thermal Hall conductivity $I^{\epsilon}_{2}$ for magnons up to second order, expressed as follows
\begin{equation}\label{nonlinear_1}
\begin{split}
    I_{2a}^{s}  
        & \approx \frac{\varepsilon_{abc}} { V\hbar^{2} T}  \sum_{n,k} c_{1}(\rho) \langle S^{z}_{n}\rangle{\partial_{k_{b}} \left[ {\Omega^{c}_{n,k}{\epsilon}_{n,k}} \right]}, \\
        I_{2a}^{\epsilon}  
        & \approx \frac{2\varepsilon_{abc}} { V\hbar^{2} T}  \sum_{n,k} c_{1}(\rho) \epsilon_{n,k}{\partial_{k_{b}} \left[ {\Omega^{c}_{n,k}{\epsilon}_{n,k}} \right]}, 
\end{split}
\end{equation}
where $V$ is volume, $\varepsilon_{abc}$ is Levi-Civita symbol with $abc=xyz$ and $c_{1}(\rho) = \int^{\rho}_{0} \mbox{log}(1+\rho^{-1}) d\rho $ with $\rho$ being Boltzmann distributions for bosons. In 2D cases, $c=z$, indicating that the nonlinear transverse current can emerge along $x$-axis when applying $\boldsymbol{\nabla}T$ along $y$-axis as shown in Fig.\,\ref{fig1}(a). Obviously, $I^{s}_{2}$ and $I^{\epsilon}_{2}$ both depend on ${\partial_{k_{y}}} \left[ {\Omega_{n,k}{\epsilon}_{n,k}} \right] = \left[\Omega_{n,k} {v^{\epsilon}_{y}} + {\epsilon}_{n,k} {v^{\Omega}_{y}} \right]$, analogous to the Berry curvature dipole\,(BCD) defined in fermions\,\cite{sodemann2015quantum}, which is referred to as the extended BCD of magnons\,\cite{kondo_prreserach_2022_4, mukherjee_PRB_2023_107, varshney2023intrinsic}. In comparison, the $I^{s}_{2}$ is associated with the expected value of $\langle S^{z}_{n}\rangle$, while $I^{\epsilon}_{2}$ pertains to the $\epsilon_{n}$ of magnons.

\begin{figure} 
\includegraphics[width=0.475\textwidth]{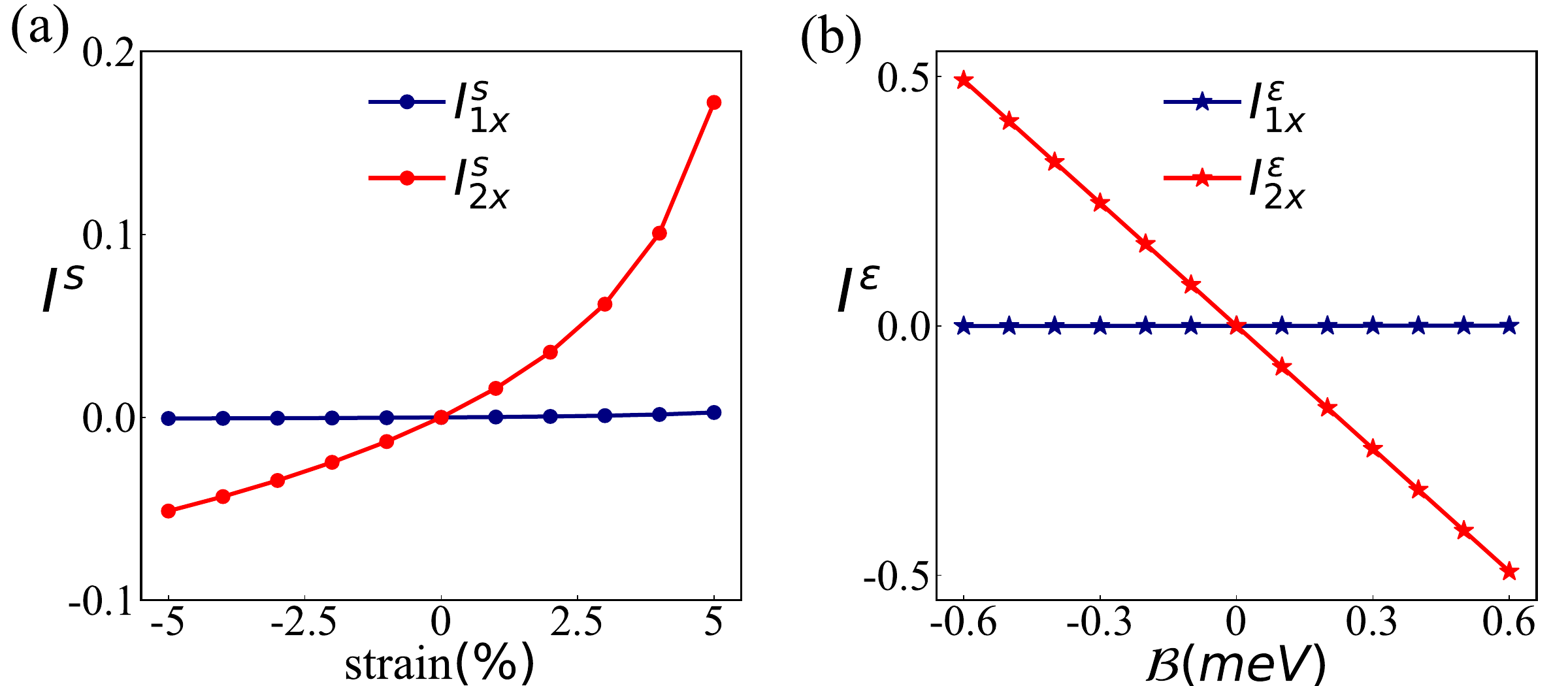}% Here is how to import EPS art
\caption{\label{fig3}\textcolor{red}{The dependence of linear\,($I_{1x}$) and nonlinear\,($I_{2x}$) Hall conductivity of magnons on the external field. (a) $I^{s}_{1x}$ and $I^{s}_{2x}$ as function of uniaxial strain along $x$-axis at a given temperature. (b) $I^{\epsilon}_{2x}$ and $I^{\epsilon}_{2x}$ as a function of ${\cal B}$. Here the temperature is fixed at $T/{\cal J}_{1}$\,$=$\,$0.6$, with both $k_{B}$ and $\hbar$ set to unity. }}
\end{figure}

As shown Fig.\,\ref{fig2}(c), the extended BCD distribution of $\alpha$ mode is even with respect to $\boldsymbol{k}$. However, as the ${\cal C}_{3z}$ symmetry can transform $k_{m}$ to $k_{\mu}$ and $k_{\nu}$ in $k$-space, the absence of ${\cal D}_{z}$ allows 
\begin{equation}\label{vel}
    v^{\epsilon/\Omega}_{y, k_{m}} + v^{\epsilon/\Omega}_{y, k_{\mu}}  + v^{\epsilon/\Omega}_{y, k_{\nu}}  = 0,
\end{equation} 
suggesting the integral of Eq.\,(\ref{nonlinear_1}) goes to zero. The nonzero summation of velocity of Eq.\,(\ref{vel}) can be generated by uniaxial strain or 2NN ${\cal D}_{z}$. The uniaxial strain can breaks the ${\cal C}_{3z}$ symmetry, which distribute the invariance among the NN spin exchange interactions ${\cal J}_{11}$, ${\cal J}_{12}$ and ${\cal J}_{13}$, as illustrated in Fig.\,\ref{fig3}(d). Consequently, there will emerge a nonlinear transverse magnon current, which is perpendicular to the
$\boldsymbol{\nabla}T$ and parallel to the direction of the uniaxial strain. Unlike uniaxial strain, ${\cal D}_{z}$ preserves the ${\cal C}_{3z}$ symmetry of honeycomb lattice, yet still results in a nonzero summation of Eq.(\ref{vel}) 
by introducing the imbalance between $\boldsymbol{k}$ and $\boldsymbol{-k}$ states. \textcolor{red}{Overall, in the context of honeycomb N\'{e}el order, the realization of a nonlinear Hall response further requires the breaking of ${\cal C}_{3z}$ symmetry or the presence of an asymmetric band.}

The absence of ${\cal B}$ ensures magnon modes remain in two-fold degenerate states, preventing the generation of the net transverse current while allowing for the nonlinear SNE\,\cite{kondo_prreserach_2022_4, mukherjee_PRB_2023_107}. This degeneracy can be lifted by applying ${\cal B}$, due to its coupling with the
$\left<S_{z}\right>$ of each magnon mode. As shown in Fig.\,\ref{fig2}(b), when ${\cal B}$\,$<$\,$0$, $\alpha$ mode is lower in energy than the $\beta$ mode, resulting in $\sum_{k}{\partial_{k_{y}} \left[ {\Omega^{z}_{\alpha}{\epsilon}_{\alpha}} + {\Omega^{z}_{\beta}{\epsilon}_{\beta}} \right] < 0 }$, such that reversing the ${\cal B}$ naturally flips the sign of $I^{\epsilon}_{2x}$. This phenomenon is fundamentally distinct from the known nonlinear Hall effect observed in fermions\,\cite{sodemann2015quantum, Ma_nature_2019_565} or phonons\,\cite{luo2023nonlinear, chen2023phonon}, as these are
independent of the magnetic field. Notice that the role of applied $\cal B$ is merely to lift the degeneracy of magnon modes, with its strength being much smaller than SIA, thereby preserving the advantage of AFM magnons with THz frequency at $\Gamma$-point \,\cite{baltz2018antiferromagnetic, gomonay2018antiferromagnetic, li2020spin}. 

Above theoretical proposal can be readily verified in various 2D layered honeycomb antiferromagnets, such as the transition metal phosphorus trichalcogenide $\mbox{VP}X_{3}$ with $X$\,$=$\,$\mbox{S}$,\,$\mbox{Se}$,\,$\mbox{Te}$. Their stability and magnetic ground state in the bulk phase have verified by recent experiments\,\cite{wang2018new, liu2023probing}, in which the honeycomb magnetic lattice within each layer, composed of $\mbox{V}^{+2}$ magnetic ion with $S$\,$=$\,$3/2$, exhibits an easy-axis N\'{e}el order below the N\'{e}el temperature $T_{N}$. Using energy mapping DFT calculations\,\cite{Supplemental_Materials,perdew1996generalized,blochl1994projector,liechtenstein1995density, xiang2011predicting, xiang2013magnetic}, we constructed the effective spin Hamiltonian for monolayer ${\mbox{VPTe}_{3}}$, finding ${\cal J}_{1}$\,$=$\, $12$\,$\mbox{meV}$, ${\cal D}_{z}$\, $=$\, $-0.47\,\mbox{meV}$ and ${\cal K}$\,$=$\,$-0.45$\,$\mbox{meV}$. The giant ${\cal D}_{z}$ and ${\cal K}$ mainly arise from the strong spin-orbital-coupling (SOC) effect of $\mbox{Te}$ ligands. Upon the application of a slight uniaxial strain along $x$-axis, ${\cal J}_{11}$ and ${\cal J}_{12}$ remain almost unchanged, whereas ${\cal J}_{13}$ change rapidly due to their strong spin-lattice coupling\,\cite{xiang2011predicting}. Thus, the large ${\cal D}_{z}$ and robust spin-lattice coupling enable it to generate giant MNHR. 

\textcolor{red}{We firstly examine the magnon nonlinear Hall response of monolayer ${\mbox{VPTe}_{3}}$ in the absence of ${\cal B}$. Fig.\,\ref{fig3}(a) shows the dependence of magnon nonlinear spin Nernst conductivity $I^{s}_{2}$ on the uniaxial strain along $x$-axis, demonstrating that a finite perturbation can induce a giant magnon nonlinear Hall response. Notably, at $1\%$ strain, the magnitude of $I^{s}_{2x}$ is about $\mbox{10}$ times greater than the contribution that at zero strain, where the response is solely attributed to the ${\cal D}_{z}$. This indicates that, compared to the intrinsic SOC, the strain effect has a more pronounced impact on the nonlinear Hall response in 2D antiferromagnets, ensuring the similar effect can be observed in ${\mbox{VPS}_{3}}$ with weak SOC\,\cite{Supplemental_Materials}. Since the extended BCD is odd under ${\cal P}$, reversing the uniaxial strain can flip the sign of $I^{s}_{2x}$. The application of ${\cal B}$ lift the degenerate magnon bands, naturally leading to the nonzero nonlinear thermal Hall response. Therefore, for ${I}^{\epsilon}_{2x}$, both the strain and ${\cal B}$ can flip it sign. In contrast to the weak response of the linear Hall response to external fields, as illustrated in Fig.\,\ref{fig3}, the nonlinear Hall response exhibits a significantly enhanced sensitivity, holding great potential for practical applications in antiferromagnetic insulators.}

\textcolor{red}{To verify the stability of magnons and resulting nonlinear Hall response under temperature variations, we include the magnon-magnon interactions into our analysis. As shown in Fig.\,S6\,\cite{Supplemental_Materials}, the four-magnons interactions leads to enhanced magnon excitability at elevated temperatures\,\cite{rezende2019introduction, mook_PRX_2021_11, sourounis2024impact}, while preserving the overall symmetry. Consequently, the nonlinear Hall response exhibits only minor corrections with temperature, and its qualitative behavior remains unchanged.}

\textcolor{red}{Building on the feature in monolayer honeycomb antiferromagnet with N\'{e}el order, breaking ${\cal P}$ but preserving ${\cal PT}$, we identify the possible layer magnetic point groups that support magneon nonlinear Hall response, as summarized in Tab.\,\ref{tab_1}. Notably, in multilayer magnets, the presence of ${\cal P}$ symmetry depends sensitively on the interlayer coupling and stacking configuration. This suggests that the emergence of the magnon nonlinear Hall effect can exhibit a pronounced layer dependence.}

\begin{figure}
\includegraphics[width=0.475\textwidth]{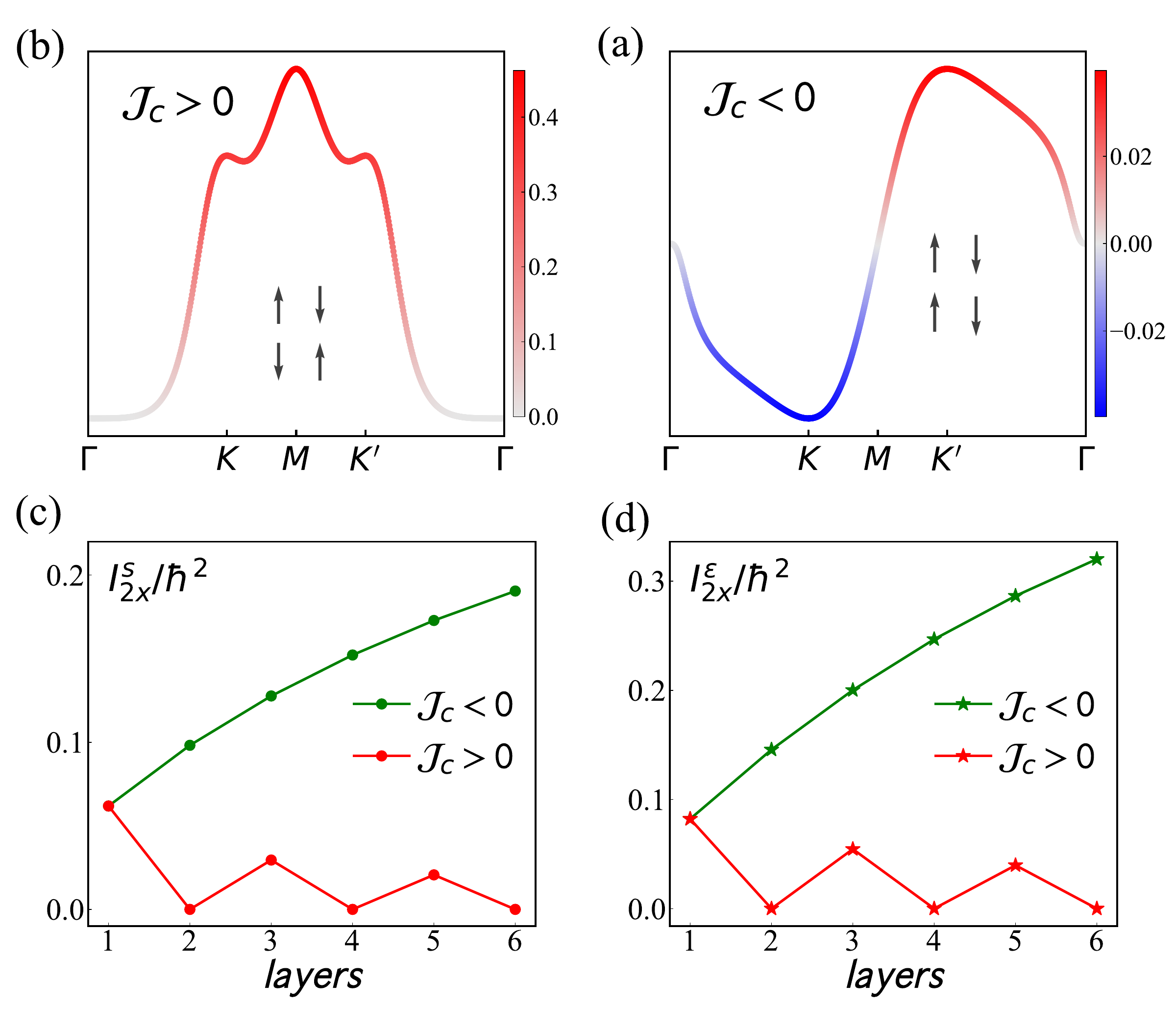}% Here is how to import EPS art
\caption{\label{fig4} \textcolor{red}{The $\boldsymbol{\Omega}$ of minor magnon band in $b$AFM, (a) with ${\cal J}_{c}$\,$<$\,$0$ and (b) ${\cal J}_{c}$\,$>$\,$0$. The dependence of $I^{s}_{2x}$ (c) and $I^{\epsilon}_{2x}$ (d) with layer numbers in honeycomb antiferromagnets with FM and AFM interlayer coupling. For $I^{\epsilon}_{2x}$, $\mbox{3}\%$ strain and ${\cal B}$\,$=$\,$0.4$\,$\mbox{meV}$ are employed. }}
\end{figure}

\textit{Layer dependence.} For bilayer honeycomb antferromagnet with AFM interlayer coupling , the ${\cal P}$ symmetry is preserved. The spin Hamiltonian of this bilayer can be read as ${\cal \hat{H}}_{b}$\,$=$\,${\cal \hat{H}}_{m}$\,$+$\,${\cal \hat{H}}_{c}$, where ${\cal H}_{c}$\,$=$\,${\cal J}_{c}\sum{{\cal S}_{i}{\cal S}_{j}}$ with ${\cal J}_{c}$\,$>$\,$0$. The momentum-space magnon Hamiltonian of this bilayer is shown in Sec.I in SM\,\cite{Supplemental_Materials}, and the corresponding eigenvalues are derived as
\begin{equation}\label{bafm_e}
\begin{split}
  \epsilon^{1}_{\pm} = &\pm {\cal B} + \sqrt{\epsilon^{2}_{c} + \epsilon^{2}_{0}-2\sqrt{\lambda^{2}{f_{k}}^{2} - |\gamma_{k}|^{2} \epsilon^{2}_{c}}} \\ 
  \epsilon^{2}_{\pm} = & \pm {\cal B} + \sqrt{\epsilon^{2}_{c} + \epsilon^{2}_{0}+2\sqrt{\lambda^{2}{f_{k}}^{2} - |\gamma_{k}|^{2} \epsilon^{2}_{c}}} 
\end{split}
\end{equation}
with ${\epsilon^{2}_{c}}=f^{2}_{k}-{\cal J}^{2}_{c}$. The eigenvectors are functions of ${\cal D}_{z}{f_{k}}$, ${\cal J}_{c}$ and $\gamma_{k}$, but are independent under ${\cal B}$. Since the ${\cal PT}^{\prime}$ symmetry of magnon Hamiltonian is preserved in the absence of ${\cal D}_{z}$, the ${\boldsymbol{\Omega}}(k)$\,$=$\,$0$ under this condition. The emergence of ${\cal D}_{z}$ can break the Dirac-points of magnon bands at $\mbox{K}$-points, as shown in Fig.\,S3\,\cite{Supplemental_Materials}, while the two-fold degenerate state remains unchanged, with each state characterized by the $\langle S^{z} \rangle$\,\cite{Supplemental_Materials}. In this scenario, the $\boldsymbol{\Omega}(k)$ of each magnon band exhibits monopole type located at the $\mbox{M}$-points as illustrated in Fig.\,\ref{fig4}(a), which is opposite to another degenerate mode. This even parity
$\boldsymbol{\Omega}(k)$ ensures that each magnon mode carries nonzero linear transverse current with respect to $\boldsymbol{\nabla}T$. Due to the even parity of $\boldsymbol{\Omega}(k)$, the corresponding BCD is odd with respect to $k$, thereby forbidding the the magnon nonlinear Hall response. This result remains robust against uniaxial strain and can be generalized to arbitrary even-layers.

\textcolor{red}{Conversely, as shown in Fig.\,\ref{fig4}(b), the ${\cal P}$ symmetry is broken when interlayer coupling becomes ferromagnetic\,(FM), i.e., ${\cal J}_{c}$\,$<$\,$0$. The two-fold degeneracy of the magnon bands is still preserved in this bilayer, and both the minor band with its associated Berry curvature remains the same as in the monolayer. The major magnon band and Berry curvature are simply shifted upward due to the interlayer coupling, expressed as\,:
\begin{equation}\label{eigen_AFM}
\begin{split}
     & {{\epsilon}}_{\pm} = \sqrt{ \lambda_{b} - |\gamma_{k}|^{2}} + f_{k} \pm {\cal B},  \\
     & \boldsymbol{\Omega}^{z}_{\pm} = \pm \frac{\lambda_{b}}{2\epsilon_{b}^{3}} \left (\boldsymbol{\nabla}{\mbox{Re}\gamma_{k}} \times \boldsymbol{\nabla}{\mbox{Im}\gamma_{k}}  \right), 
\end{split}
\end{equation}
where $\lambda_{b}$\,$=$\,$\lambda$\,$-$\,2${\cal J}_{c}$ and ${\epsilon}_{b}$\,$=$\,$\sqrt{\lambda^{2}_{b} - |\gamma_{k}|^{2}}$. Fig.\,\ref{fig4}(b) shows the ${\Omega}$ of major band, which also exhibits odd parity similar to minor band. However, its strength is suppressed by interlayer coupling, thereby reducing contribution of this band to the nonlinear response. The layer dependence of magnon nonlinear Hall response in the multilayers with ferromagnetic ${\cal J}_{c}$ is shown in Fig.\,\ref{fig4}(c)–(d), demonstrating a positive correlation.}

\textcolor{red}{For comparison, we calculate the layer-dependent nonlinear Hall effect in the case of AFM interlayer coupling. As shown in Fig.\,\ref{fig4}(c)-(d), in addition the pronounced odd-even layer dependence constrained by the symmetry, the strength of the nonlinear Hall effect gradually decreases with increasing layer number. This feature offers an effective route to characterize the nature and strength of magnetic interlayer coupling in few-layers AFM insulators.} 

\textit{Discussion and summary.} \textcolor{red}{In addition to the honeycomb antiferromagnets with N\'{e}el order, magnon nonlinear responses can also be realized in layered magnets of which monolayer adopts ${\cal P}$ symmetry, by means of stacking engineering. For example, in bilayer $\mbox{CrI}_{3}$, FM order within each layer preserves ${\cal P}$ symmetry, while the AFM stacking configuration breaks this symmetry\,\cite{sun2019giant, sivadas2018stacking}. Moreover, in vdW multilayers, the magnetic coupling between layers can be tuned by applying a finite external field\,\cite{huang2018electrical, jiang2018controlling, ni2025nonvolatile}, providing an effective means for external-field control of the nonlinear response.}
This magnetic-symmetry-constrained rule for nonlinear Hall effect of magnons contrasts with the linear magnon Hall effect, which requires only that ${\cal C}_{2}{\cal T}$ to be broken by ${\cal D}_{z}$, independent of the magnetic order\,\cite{onose_science_2010_329,matsumoto2011theoretical, katsura_prl_2010_104, zhang_PRL_2021_127,mook_PRX_2021_11,owerre_NJP_2016_28, zhang_thermal_Physreport_2024_1070, matsumoto_PRB_2014_89}. Therefore, magnon nonlinear Hall effect is particularly advantageous for probing the magnetic structures of antiferromagnets, such as detecting the spin order within each layer and characterizing the interlayer coupling. 

A similar nonlinear transport effect has also been recently observed in electrons within ${\cal PT}$ antiferromagnets, as seen in even-layer $\mbox{Mn}\mbox{Bi}_{2}\mbox{Te}_{4}$\,\cite{wang2023quantum, gao2023quantum} and monolayer $\mbox{MnS}$\,\cite{wang2023intrinsic}. This newly discovered nonlinear Hall effect of electrons is induced by the quantum metric, real part of quantum geometry tensor of electron Bloch wavefunctions\,\cite{wang2023quantum, gao2023quantum, wang2023intrinsic}. In addition, the nonlinear Hall effect of magnons proposed here exhibits several distinctions from the electrons. Firstly, nonlinear Hall effect in electrons require a nonzero fermi-surface or a narrow band gap, but magnons can overcome this limitation. Secondly, nonlinear Hall effect of electrons is independent of external fields, whereas magnons is highly tunable by the ${\cal B}$ or strain.   

In summary, we have theoretically investigated the magnon nonlinear Hall effect in 2D vdW AFM insulators, revealing their remarkable tunability and intriguing magnetic-dependent characteristic. Our findings not only shed light on potential applications of AFM spintronics, but also provide an entirely new approach for detecting the magnetic order in 2D vdW magnetic insulators.

We acknowledge useful discussions with Prof.\,Hongjun Xiang, Prof.\,Zhijun Jiang, Prof.\,Arijit Kundu, Dr.\,Hou Tao and Dr.\,Rohit Mukherjee. This work at Nanyang Technological University was financially supported by grants from the National Research Foundation, Singapore under its Fellowship Award (NRF-NRFF13-2021-0010), the Nanyang Technological University startup grant (NTUSUG). Y.\,Jin thanks to the Guangdong Provincial Quantum Science Strategic Initiative (Grant No. GDZX2401002), the Natural Science Foundation of China (Grant No. 12404181). The Singapore National Super
Computing Centre (NSCC) is also acknowledged.  

\appendix
%\begin{verbatim}
%\end{verbatim}

\nocite{*}
%\nobalance
%\bibliographystyle{plain}
\bibliographystyle{abbrv}
\bibliographystyle{elsarticle-num-names}
\bibliographystyle{apsrev4-2}
\bibliography{Main}% Produces the bibliography via BibTeX.

\end{document}